\input harvmac

\def \Tr {{\rm Tr}}

\def\a{\alpha}

\def \a {\alpha}
\def \b {\beta}
\def \ov {\over}

\def \ov {\over }
\def \four{{\textstyle{1\over 4}}}

\def \lr { \lref}

\def \four{{\textstyle{1\over 4}}}

\def\a{\alpha} 
\def \b {\beta}
 
\def \ov {\over}

\def \third { { \textstyle{ 1 \ov 3} } }

\def\np {  {\it Nucl. Phys.} }
\def \pl { {\it  Phys. Lett.} }

\def \pr  { {\it Phys. Rev.} }

\lr\KT{I.R. Klebanov and A.A. Tseytlin, ``D-Branes and
Dual Gauge Theories in Type 0 Strings,''
\np {\bf B546} (1999) 155, {\tt hep-th/9811035}. }

\lr\KTc{I.R. Klebanov and A.A. Tseytlin, ``Non-supersymmemtric CFT
from Type 0 String Theory,'' 
{\it J. High Energy Phys.}  {\bf 9903} (1999)  015, {\tt hep-th/9901101}. 
}

\lr\GA{ M.R. Garousi, 
``String Scattering from D-branes in Type 0 Theories'',
{{\tt hep-th/9901085}}.}

\lr\ber{M. Bershadsky, Z. Kakushadze and
C. Vafa, 
``String expansion as large N expansion of gauge theories",
\np {\bf B523} (1998) 59, {\tt hep-th/9803076};
M. Bershadsky and A. Johansen,
``Large N limit of orbifold field theories,''
{\it Nucl. Phys.} {\bf B536} (1998) 141,
{\tt hep-th/9803249}.}

\lr\DM{M. Douglas and G. Moore, ``D-branes, quivers, and ALE
instantons,'' {{\tt hep-th/9603167}}.
}

\lr \DH { L. Dixon and J. Harvey, ``String theories in ten dimensions without space-time supersymmetry",  
{\it Nucl. Phys.} {\bf B274} (1986) 93;
 N. Seiberg and E. Witten,
``Spin structures in string theory", 
{\it Nucl. Phys.} {\bf B276} (1986) 272;
C. Thorn, unpublished.}

\lr\JM {J. Minahan, ``Glueball Mass Spectra and Other Issues for 
Supergravity Duals of QCD Models,'' {\tt hep-th/9811156}.  }

\lr\JMnew {J. Minahan, ``Asymptotic Freedom and Confinement from Type 0
String Theory,'' {\tt hep-th/9902074}.  }

\lr \ferr{G. Ferretti and D. Martelli, 
``On the construction of gauge theories from non critical type 0 strings,"
{\tt hep-th/9811208}. }

\lr\Sasha{A.M. Polyakov, ``String theory and quark confinement,''
{\it Nucl. Phys. B (Proc. Suppl.)} {\bf 68} (1998) 1, {{\tt hep-th/9711002}}. }

\lr\berg{O. Bergman and M. Gaberdiel, ``A Non-supersymmetric Open
String Theory and S-Duality,'' \np {\bf B499} (1997) 183,
{\tt hep-th/9701137}. }

\lr\bg{O. Bergman and M. Gaberdiel, ``Dualities of Type 0 Strings,''
{\tt hep-th/9906055}. }

\lr  \kleb{
I.R. Klebanov, ``World volume approach to absorption by nondilatonic branes,''
  {\it Nucl. Phys.} {\bf B496} (1997) 231,
  {{\tt hep-th/9702076}}; 
S.S. Gubser, I.R. Klebanov, and A.A. Tseytlin, ``String theory and classical
 absorption by three-branes,'' {\it Nucl. Phys.} {\bf B499} (1997) 217,
  {{\tt hep-th/9703040}}.}

\lr  \gkThree{
S.S. Gubser and I.R. Klebanov, ``Absorption by branes and Schwinger terms in
  the world volume theory,'' {\it Phys. Lett.} {\bf B413} (1997) 41,
  {{\tt hep-th/9708005}}.}

\lr  \jthroat{
J.~Maldacena, ``The Large N limit of superconformal field theories and
  supergravity,'' {\it Adv. Theor. Math. Phys.} {\bf 2} (1998) 231, 
{{\tt
  hep-th/9711200}}.}

\lr  \US{
S.S. Gubser, I.R. Klebanov, and A.M. Polyakov, ``Gauge theory correlators
  from noncritical string theory,'' {\it Phys. Lett.} {\bf B428} (1998)
105, {{\tt hep-th/9802109}}.}

\lr  \EW{
E.~Witten, ``Anti-de Sitter space and holography,''
 {\it Adv. Theor. Math. Phys.} {\bf 2} (1998) 253, 
 {{\tt hep-th/9802150}}.}

\lr\Kehag{
A. Kehagias, ``New Type IIB Vacua and Their F-Theory Interpretation,''
{{\tt hep-th/9805131}}.
}

\lr  \AP{
A.M. Polyakov, ``The Wall of the Cave,''
{\tt hep-th/9809057.}}

\lr  \brane{
J.~Polchinski, ``Dirichlet Branes and Ramond-Ramond charges,'' {\it Phys. Rev.
  Lett.} {\bf 75} (1995) 4724,
{{\tt hep-th/9510017}}. }

\lr  \Jbook{
J. Polchinski, ``String Theory,'' vol. 2, Cambridge University Press,
1998.}

      \lr\TASI{
J. Polchinski, ``TASI Lectures on D-Branes,''
{\tt hep-th/9611050.}  }

\lr  \Witten{
E.~Witten, ``Bound states of strings and p-branes,'' {\it Nucl. Phys.} {\bf
  B460} (1996) 335, {{\tt
  hep-th/9510135}}.  }

\lr\GTone{
G. `t Hooft, ``On the Convergence of Planar Diagram Expansions,''
{\it Comm. Math. Phys.} {\bf 86} (1982) 449.}

\lr\KW{I.R. Klebanov and E. Witten, ``Superconformal field theory on
threebranes at a Calabi-Yau singularity,''
\np {\bf B536} (1998) 199, {\tt hep-th/9807080};
S.S. Gubser, ``Einstein manifolds and conformal field theories,''
{\tt hep-th/9807164}.}

\lr\KWnew{I.R. Klebanov and E. Witten, ``AdS/CFT Correspondence and
Symmetry Breaking,'' {\tt hep-th/9905104}.}

\lr\BIPZ{E. Brezin, C. Itzykson, G. Parisi and J. Zuber,
{\it Comm. Math. Phys.} {\bf 59} (1978) 35.}

 \lr\jones{D.R.T. Jones,
``Asymptotic behavior  of supersymmetric Yang-Mills theories
in the two-loop approximation," {\it Nucl. Phys.} {\bf  B87}
 (1975) 127; ``Charge renormalization 
in a supersymmetric Yang-Mills theory,"
{\it  Phys. Lett.}  {\bf B72} (1977) 199; 
E. Poggio and H. Pendleton,  
``Vanishing of charge renormalization 
and anomalies in a supersymmetric gauge theory,''
 {\it  Phys. Lett.}  {\bf B72}  (1977) 200.}

\lr\MV{ 
M.E. Machacek and M.T. Vaughn, 
``Two-loop renormalization group equations in a general 
quantum field theory I: Wave function renormalization,"
{\it Nucl. Phys.} {\bf B222} (1983) 83;
``II. Yukawa couplings'', \np  {\bf B236} (1983) 221;
``III. Scalar quartic  couplings'', \np  {\bf B249} (1985) 70.
}

\lr\KTnew{
I.R. Klebanov and A.A. Tseytlin, ``Asymptotic Freedom and
Infrared Behavior in the Type 0 String Approach to Gauge Theory,'' 
\np {\bf B547} (1999) 143, {\tt hep-th/9812089}. }

\lr \KS  {S.~Kachru and E.~Silverstein, ``4d conformal field theories
and strings on orbifolds,''  {\it Phys. Rev. Lett. } {\bf 80} (1998)
4855, 
{{\tt hep-th/9802183}}.}
\lr\LNV {A.~Lawrence, N.~Nekrasov and C.~Vafa, ``On conformal field
theories in four dimensions,'' {\it Nucl. Phys.} {\bf B533} (1998) 199, 
{{\tt hep-th/9803015}}.}

\lr \FR { P. Frampton, ``ADS/CFT String Duality and Conformal Gauge
Theories,'' {\tt hep-th/9812117};
P. Frampton and C. Vafa, ``Conformal Approach to Particle
Phenomenology,'' {\tt hep-th/9903226}. }

\lr \DB  {J. Blum and   K.  Dienes, ``Strong / Weak Coupling Duality
Relations for
Non Supersymmetric String Theories",  {\it Nucl. Phys.} {\bf  B516}  
(1998) 83, 
{\tt hep-th/9707160.} } 

\lr\bisa{
M. Bianchi and  A. Sagnotti,
``On the Systematics of Open String Theories",
{\it Phys. Lett.} {\bf B247}  (1990) 517;
A. Sagnotti, ``Some Properties of Open - String Theories", 
{\tt hep-th/9509080}; ``Surprises in Open-String Perturbation Theory", 
 {\it Nucl.Phys.Proc.Suppl.} {\bf  B56}  (1997) 332, 
{\tt hep-th/9702093}.}

\lr\BF{P. Breitenlohner and D.Z. Freedman,
``Stability in gauged extended supergravity'',
{\it Ann. Phys.} {\bf 144} (1982) 249.}

\lr\DS{
M. Dine, P. Huet and N. Seiberg, 
``Large and small radius in string theory'',
{\it Nucl. Phys.}   {\bf B322} (1989) 301;
 R. Rohm,
``Spontaneous supersymmetry breaking in supersymmetric field theories'',
 \np {\bf B237}  (1984) 553.}

\lr\AW{J.J. Atick  and E. Witten, 
``The Hagedorn transition and the number of degrees of freedom 
of string theory'', 
\np {\bf B310} (1988) 291.}

\lr\newWit{
E. Witten, ``Anti-de Sitter space, thermal
phase transition, and confinement in gauge theories,''
{\it Adv. Theor. Math. Phys.} {\bf 2} (1998) 505, {\tt hep-th/9803131.}
}

\lr\JO{I. Jack and H. Osborn, ``General background field calculations with fermion fields", 
\np {\bf B249} (1985) 472.}
\lr \OO{T.P. Cheng, E. Eichten and L.-F. Li, 
``Higgs phenomena in asymptotically free gauge theories,"
\pr {\bf D9} (1974) 2259.}

\lr\NS{N. Nekrasov and S. Shatashvili,
``On non-supersymmetric CFT in four dimension,''
{\tt hep-th/9902110.}
}

\lr\BCR{M. Bill\' o, B. Craps and  F. Roose, ``On D-branes in 
Type 0 String Theory,'' {\tt hep-th/9902196.}
}

\lr\TZ{K. Zarembo, ``Coleman-Weinberg Mechanism and Interaction of
D3-branes in Type 0 String Theory, {\tt hep-th/9901106};
A. Tseytlin and K. Zarembo, ``Effective Potential in
Non-supersymmetric $SU(N)\times SU(N)$ Gague Theory and
Interactions of Type 0 D3-branes,''
{\tt hep-th/9902095.}
}

\lr\tern{C. Csaki, W. Skiba and J. Terning,
``Beta Functions of Orbifold Theories and the Hierarchy Problem,''
{\tt hep-th/9906057.}}

\lr\Costa{M. Costa,
``Intersecting D-branes and black holes in type 0 string theory,''
{\tt hep-th/9903128}.}

\lr\Oz{M. Alishahiha, A. Brandhuber and Y. Oz,
``Branes at Singularities in Type 0 String Theory,''
{\tt hep-th/9903186.}
}

\lr\MP{D.R. Morrison and M.R. Plesser,
``Non-Spherical Horizons, I,'' hep-th/9810201.}

\lr\AK{A. Armoni and B. Kol, ``Non-supersymmetric Large $N$
Gauge Theories from Type 0 Brane Configurations,''
{\tt hep-th/9906081.}
}

\lr\Blum{R. Blumenhagen, A. Font and D. Lust,
``Non-Supersymmetric Gauge Theories from 
D-Branes in Type 0 String Theory,''
{\tt hep-th/9906101.}}

\lr\Guk{S. Gukov, ``Comments on N=2 AdS Orbifolds,''
\pl {\bf B439} (1998) 23, {\tt hep-th/9906081.}}


\baselineskip8pt
\Title{\vbox
{\baselineskip 6pt
{\hbox {PUPT-1874}}
{\hbox{hep-th/9906220}} 
{\hbox{   }}
}}
{\vbox{\vskip -30 true pt
\centerline {Tachyon Stabilization in the AdS/CFT Correspondence}
\medskip
\vskip4pt }}
\vskip -20 true pt 
\centerline{ Igor R. Klebanov}
\smallskip\bigskip
\centerline{Joseph Henry Laboratories} 
\smallskip
\centerline{Princeton University} 
\smallskip
\centerline{Princeton, New Jersey 08544} 
\smallskip
\centerline{USA}
\bigskip\bigskip
\centerline {\bf Abstract}
\baselineskip12pt
\noindent
\medskip
We consider duality between type 0B string theory on $AdS_5\times S^5$
and the planar CFT on $N$ electric D3-branes coincident with $N$
magnetic D3-branes. It has been argued that this theory is
stable up to a critical value of the `t Hooft coupling but is unstable
beyond that point. We suggest that from the gauge theory point
of view the development of instability is associated with singularity
in the dimension of the operator corresponding to the tachyon field
via the AdS/CFT map. Such singularities are common in large $N$ theories
because summation over planar graphs typically has a finite radius of
convergence. Hence we expect transitions between stability and
instability for string theories in AdS backgrounds that are dual
to certain large $N$ gauge theories: if there are tachyons for large
AdS radius then they may be stabilized by reducing the radius below
a critical value of order the string scale.

\bigskip
 
\Date {June 1999}

\noblackbox \baselineskip 15pt plus 2pt minus 2pt 

\newsec{Introduction}

It has been proposed \refs{\KT,\JM,\KTnew,\KTc} 
that certain 4-dimensional non-supersymmetric 
large $N$ gauge theories are dual to backgrounds of
the type 0 string theory \DH.  
This work was motivated by the recently discovered relations between
type IIB strings and superconformal gauge theories on $N$ coincident
D3-branes \refs{\kleb,\gkThree,\jthroat,\US,\EW} as well as 
by Polyakov's suggestion \refs{\AP} (building on his earlier work
\Sasha)
that the type 0 string theory is a natural
setting for extending this duality to {\it non}-supersymmetric
gauge theories. Since then a number of further studies of D-branes
in type 0 theories \refs{\GA,\BCR,\TZ}
and the associated dualities 
\refs{\ferr,\JMnew,\NS,\Costa,\Oz,\bg,\AK,\Blum} have appeared.

It is well known that taking the low
energy limit on a stack of $N$ D3-branes in flat space
and comparing it to the corresponding limit of the
3-brane classical solution suggests that the
${\cal N}=4$ supersymmetric $SU(N)$ gauge theory is dual to
type IIB strings on $AdS_5\times S^5$ \refs{\jthroat,\US,\EW}.
Generalizations of this duality to other 4-dimensional gauge theories
follow if the stack of D3-branes is placed on a transverse space $Y_6$
which is a cone over an Einstein manifold $X_5$
\refs{\DM,\KS,\LNV,\Kehag,\KW,\MP,\KWnew}.
Then the the CFT on the
D3-branes is dual to type IIB string theory on
$AdS_5\times X_5$. 
An interesting set of dualities of this sort is
found if $Y_6$ is an orbifold $R^6/\Gamma$ where $\Gamma$ is
a discrete subgroup of $SO(6)$ \refs{\KS,\LNV}. The corresponding
theories on the branes are conformal if the orbifold group acts
appropriately on the gauge indices \DM.  
Duality between such CFT's and type IIB backgrounds of the
form $AdS_5\times S^5/\Gamma$ is thus an orbifold of the original
duality between the ${\cal N}=4$ theory and the $AdS_5\times S^5$
background. Remarkably, a large
class of such CFT's is non-supersymmetric in the large $N$ (planar)
limit \refs{\KS,\LNV}. A phenomenological application of
non-supersymmetric CFT's was recently proposed in \FR\ but later
criticized in \tern.

Alternatively, one may construct gauge theory/string dualities
by stacking D3-branes in string models other than type IIB.
Type 0 theories are attractive in this regard 
but they have a well-known
and seemingly fatal flaw in that their spectrum contains a tachyon
of $m^2=-2/\alpha'$ which is not removed by the non-chiral GSO
projection $(-1)^{F+\tilde F}=1$. In \KT\ it was suggested,
however, that the tachyon may be stabilized in appropriate type 0
backgrounds which are relevant to gauge theory applications.

The type 0 models have a doubled
set of RR fields compared to their type II cousins, hence they
also possess twice as many D-branes \KT. For example, since type 0B
spectrum has an unrestricted 4-form gauge potential, there are
two types of D3-branes: those that couple electrically to this
gauge potential, and those that couple magnetically.
Very importantly, the weakly coupled spectrum of open strings on
type 0 D-branes does not contain tachyons after the GSO
projection $(-1)^{F_{open}}=1$ 
is implemented \refs{\KT,\AP,\berg}. Thus, gauge theories living on
such D-branes do not have obvious instabilities. This suggests via
the gauge field/string duality that the bulk tachyon instability
of type 0 theory may be cured as well \refs{\KT,\KTc}.

Let us review an argument for the tachyon stabilization
in the simplest setting,
which is the stack of $N$ electric and $N$ magnetic D3-branes \KTc.
For such a stack the net tachyon tadpole cancels so that there
exists a classical solution with $T=0$. In fact, since the stack couples
to the selfdual part of the 5-form field strength, the type 0B
3-brane classical
solution is identical to the type IIB one. Taking the throat
limit suggests that the low-energy field theory on
$N$ electric and $N$ magnetic D3-branes is dual to the $AdS_5\times S^5$
background of type 0B theory and is therefore conformal in the planar
limit \KTc. This theory is the $U(N)\times U(N)$
gauge theory coupled to 6 adjoint scalars of the first $U(N)$,
6 adjoint scalars of the second $U(N)$, and fermions in the
bifundamental representations -- 4 Weyl fermions in the
$({\bf N}, \overline {\bf N})$ and 4 Weyl fermions in the
$(\overline {\bf N}, {\bf N})$ (the $U(1)$ factors decouple in
the infrared). While such a theory does not
correspond to placing type IIB D3-branes at an orbifold point, it is
nevertheless a $Z_2$ projection of the ${\cal N}=4$ $U(2N)$ gauge
theory \KTc. The $Z_2$ is generated by 
$(-1)^{F_s}$, where $F_s$ is the fermion number,
together with conjugation by
$\pmatrix{ I&  0\cr  0 & -I\cr}$ 
where $I$ is the $N \times N$ identity matrix. 
This is related to the fact that
type 0 string theories may be viewed
as $(-1)^{F_s}$ orbifolds of the corresponding type II theories \DH.  
In \NS\ it was pointed out that,
since $(-1)^{F_s}$ is an element of the center of the $SU(4)$
R-symmetry, this $Z_2$ projection of the ${\cal N}=4$ theory
belongs to the class studied in \refs{\KS,\LNV}. 

The fact that the
non-supersymmetric AdS/CFT duality considered in \KTc\ is a
$Z_2$ quotient of the ${\cal N}=4$ duality \refs{\KTc,\NS} lends
additional credence to its validity. In particular, general arguments
presented in \ber\ guarantee that the field theory is conformal in the planar
limit and that planar correlators of untwisted vertex operators
coincide with those of the ${\cal N}=4$ theory.
Since this CFT does not appear to have any instabilities at weak
coupling, it was argued in \KTc\ 
that type 0B string theory on $AdS_5\times S^5$ is 
stable for sufficiently small radius. This provides
a simple AdS/CFT argument in favor of tachyon stabilization; it is also an
example of how gauge theory may be used to make predictions about
the string theory dual to it.

We should note that some important aspects of 
this particular $Z_2$ quotient
are related to the twisted sector of the orbifold and are
not covered by the analysis in \ber. To analyze some of the
physics it will be important that this theory is dual to type 0B
rather than type IIB background.
Therefore, in the usual gravity limit where the radius 
of $AdS_5\times S^5$ becomes large in string units
and space becomes locally flat,
this background is unstable \KTc. From the CFT point of view it
means that the limit of infinite `t Hooft coupling, which is commonly
discussed in the AdS/CFT context, does not make sense.
The AdS/CFT correspondence applied to this particular
case implies that, as $\lambda = g_{YM}^2 N$ is increased,
a transition should happen at a critical value $\lambda_c$ \KTc.
In this paper we address the origin of this transition from the
field theory point of view and argue that it is due to a singularity
of a sum over planar diagrams. In non-supersymmetric large $N$
field theories such singularities are expected to occur on general
grounds \GTone.\foot{In view of this fact, which is also supported
by evidence from exactly solvable large $N$ models \BIPZ, 
it may seem surprising that certain non-supersymmetric
planar CFT's appear to make sense at infinite `t Hooft coupling \KS.
A possible explanation is that in these theories $\lambda_c$
is negative. }
We identify the field theory quantity which we expect to become
singular at $\lambda_c$: it is the dimension of the operator which
corresponds to the tachyon field via the AdS/CFT map. Using the methods
of \kleb\ we show that this operator is
\eqn\tachyon{
{\cal O}_T= {1\over 4} \Tr F_{\a\b}^2 - {1\over 4} \Tr G_{\a\b}^2 
+{1\over 2} \Tr (D_\alpha X^i)^2 - {1\over 2}\Tr (D_\alpha Y^i)^2 + \ldots
\ ,
}
where $F_{\a\b}$ and $X^i$ are the field strength and the 6 adjoint
scalars of $SU(N)_1$ while
$G_{\a\b}$ and $Y^i$ are the corresponding objects of $SU(N)_2$.
Adding such an operator to the action creates a difference between
the gauge couplings of $SU(N)_1$ and $SU(N)_2$. In non-supersymmetric
field theory this is not expected to be an exactly marginal deformation.
By analyzing 2-loop beta functions we show that the operator
\tachyon\ indeed picks up an anomalous dimension of order $\lambda^2$.
Higher loops graphs will correct the dimension $\Delta$ as well, and
it is conceivable that $\Delta(\lambda)$ develops a singularity at
some critical value $\lambda_c > 0$ as suggested by the AdS considerations.

\newsec{Anomalous dimension of the tachyon operator}

The standard relation between scalar operator dimension in a conformal
field theory and mass in $AdS_5$ is \refs{\US,\EW,\KWnew}
\eqn\relation{ \Delta = 2 \pm \sqrt{4 + m^2 L^2}
\ .
}
This relation is definitely valid in the gravity limit where
$ L^2/\alpha' = \sqrt{2\lambda} \gg 1$. Applying this relation to the case
of type 0 tachyon, $m^2= -2/\alpha'$, we immediately find that
for large $\lambda$ $\Delta$ is complex, hence the CFT is unstable.
From the AdS point of view, the stability bound \BF\ that
mass-squared must exceed $-4/L^2$ is violated for sufficiently large $L$.
For $\lambda$ of order 1 (i.e. for AdS radius of order
the string scale) there may be various corrections to the
relation \relation. Even if the overall form of this relation
remains the same, $L^2/\alpha'$ should be treated as some unknown
function of $\lambda$, and effective $m^2$ might have the form
\eqn\mass{
m^2 = -{2\over \alpha'} + {a_1\over L^2} + 
{a_2 \alpha' \over L^4} + \ldots
}
This forces us to proceed qualitatively.

To begin with let us simply continue the gravity relation 
to all $\lambda$ without any change.
Then we find
\eqn\newrelation{ \Delta (\lambda) = 2 \pm \sqrt{4 - 2\sqrt{2\lambda} }
\ .
}
The dimension is real for sufficiently small $\lambda$, and the
unitarity bound $\Delta > 1$
forces us to pick the positive branch of the square root.
This simplified ``model'' for $\Delta(\lambda)$ implies that
$\Delta (0)=4$ which, as we will see, is the correct answer.
However, obviously \newrelation\ cannot be exact: it gives
an expansion in powers of $\sqrt \lambda$ while in the gauge theory
it has to proceed in integer powers of $\lambda$ (we will see that the leading
anomalous dimension is $O(\lambda^2)$).
We expect various corrections to 
this ``model'' such as those discussed above.
For example, tachyon interactions with the 
R-R 5-form field strength, which have the form $F_5^2 T^2$ \KT, give
rise to a positive $a_1$ in \mass. One effect of this shift
is to push $\lambda_c$ to a higher value, but we cannot rule
out the presence of other similar corrections. Our qualitative picture
is quite insensitive to details however: $\Delta(\lambda)$
is real for $\lambda < \lambda_c$ but has a branch cut for
$\lambda > \lambda_c$ which signals an instability.
In the following we will use perturbative field theory arguments to
support this scenario.

The primary question is: what is the operator in the
CFT on $N$ electric and $N$ magnetic D3-branes that couples to the
tachyon. 
The tachyon coupling to a stack of $N$ electric D3-branes is
\refs{\KT,\KTnew,\GA}
\eqn\coupnew{
- T_3 \int d^4 x \ k_e(T)\  (N + {1\over 4} \Tr F_{\a\b}^2 +
{1\over 2} \Tr (D_\a X^i)^2 + 
\ldots ) 
\ , }
where $k_e(T)= 1 + \four T + O(T^2)$ and $T_3$ is the tension of
an electric D3-brane.
The coupling to a stack of $N$ magnetic D3-branes is instead
\refs{\KT,\KTnew,\GA}
\eqn\coupnewm{
- T_3 \int d^4 x \ k_m(T)\  (N + {1\over 4} \Tr G_{\a\b}^2 +
{1\over 2} \Tr (D_\a Y^i)^2 + 
\ldots ) 
\ , }
where $k_m(T)= 1 - \four T + O(T^2)$.
Adding these two actions, we find that
the tachyon coupling to a stack 
of $N$ electric and $N$ magnetic branes is
\eqn\op{
\sim \int d^4 x\ T ({1\over 4} \Tr F_{\a\b}^2 - {1\over 4} \Tr G_{\a\b}^2
+ {1\over 2} \Tr (D_\a X^i)^2 - {1\over 2} \Tr (D_\a Y^i)^2
+ \ldots  )
}
leading to the expression \tachyon\ for the gauge invariant
operator corresponding to the tachyon field via the AdS/CFT map. 
This operator is odd under the interchange
of the two $SU(N)$'s which is related to the fact that the tachyon is
one of the twisted states from the point of view of orbifolding type
IIB string theory by $(-1)^{F_s}$. Indeed, the
tachyon belongs to the $(NS-, NS-)$ sector of type 0B theory
which is not present in the
type IIB spectrum. In other $Z_2$ orbifold theories it was also found that
twisted states couple to vertex operators odd under the interchange
of the factors of the gauge group \refs{\Guk,\KW}.
Fields from the untwisted sector, on the other hand, couple to
operators that are even under the interchange of the gauge groups:
for example, the dilaton couples to
\eqn\dilaton{
{1\over 4} \Tr F_{\a\b}^2 + {1\over 4} \Tr G_{\a\b}^2 + \ldots
\ .
}

In order to study the anomalous dimension of the operator
\tachyon\ at weak coupling, we may consider slightly unequal gauge couplings
for $SU(N)_1$ and $SU(N)_2$. Then the coupling constant for this
operator is
\eqn\tren{ T = {1\over \lambda_1} - {1\over \lambda_2}\ ,
}
where $\lambda_i= g_i^2 N$, and $i=1,2$.
The planar (leading order in $N$)
renormalization group equations to two-loop order are \KTc
\eqn\first{\mu {\partial\over \partial \mu} {1\over \lambda_1} =
a (\lambda_1 - \lambda_2) + O\left [(\lambda_1-\lambda_2)^2\right ]\ ,
}
\eqn\second{\mu {\partial\over \partial \mu} {1\over \lambda_2} =
a (\lambda_2 - \lambda_1) + O\left [(\lambda_1-\lambda_2)^2\right ]\ ,
}
and we find $a=-(2\pi)^{-4}$.
Due to the vanishing of the one-loop beta function there is no
constant term on the right-hand side of the equations. The
two-loop contributions vanish for equal couplings, as shown
explicitly in \KTc, but they no longer vanish for
$\lambda_1 \neq \lambda_2$. 

Taking the difference of the two equations, and working to first order
in $T$, we have
\eqn\third{\mu {\partial T\over \partial \mu} = {\lambda^2\over 8 \pi^4}
 T
\ .}
Solving this we have
\eqn\sol{
T= T_0 \left ({\mu\over \Lambda}\right )^{{\lambda^2\over 8 \pi^4}}
\ .
}
Since $T$ grows in the UV (for $\mu\gg \Lambda$) the associated operator
\tachyon\ is irrelevant to this order: its dimension is
\eqn\pert{\Delta(\lambda)= 4 + {\lambda^2\over 8 \pi^4}+ O(\lambda^3)
\ .
}
Based on the AdS/CFT correspondence,
we conjecture that further corrections to $\Delta(\lambda)$ will
produce a function which develops a singularity at $\lambda_c>0$.
We do not expect direct calculations of the anomalous
dimension to be feasible beyond
$O(\lambda^3)$ which may be extracted from 3-loop beta functions.
However, it might be possible to put bounds on the general
$O(\lambda^n)$ terms and prove the existence of a critical point at
$\lambda_c>0$.

To summarize this paper, we have suggested that finite radius of
convergence of planar graphs, which is a well-known phenomenon
in large $N$ field theories, has interesting implications in view of
the AdS/CFT correspondence. Namely, if there is an operator whose
dimension is real up to a critical value of the `t Hooft coupling
but becomes complex beyond that value, then the dual AdS phenomenon
is that a tachyon present in the large radius (gravity) limit is
stabilized for sufficiently small radii. It is difficult to study
such backgrounds with string-scale curvature directly, but on the dual
CFT side perturbative calculations are quite tractable.
The perturbative stability of the CFT thus provides an argument for
stabilization of tachyons (this is one of the few results to date
where field theory has been used to make new predictions about string theory),
and finiteness of the radius of convergence of planar graphs suggests
how instability may develop at sufficiently strong coupling.
We have discussed one specific example which arises in the context of
type 0 theory but we hope that these phenomena are quite general.

\bigskip
\noindent
{\bf Acknowledgements}
\bigskip
I am grateful to 
E. Silverstein, A. Tseytlin and especially J. Minahan 
for useful discussions
and comments.  This work  was supported in part by the NSF
grant PHY-9802484 and by the James S. McDonnell
Foundation Grant No. 91-48.
\vfill\eject
\listrefs
\end